\documentstyle[11pt,newpasp,psfig]{article}

\begin{document}

\title{Stellar Streams in the Solar Neighbourhood from High Resolution
$N$-Body Simulations}
\author{R. Fux}
\affil{
       Mount Stromlo Observatory, Weston Creek P.O., ACT 2611\\
       Geneva Observatory, Ch. des Maillettes 51, CH-1290 Sauverny}

\begin{abstract}
A high-resolution $N$-body simulation suggests that stellar streams in the
discs of barred galaxies are common and strongly time-dependent. The velocity
distribution of stars in the Solar neighbourhood betray many such streams,
including a stream of outward moving stars with low angular momentum.
This stream is interpreted as a signature of the Galactic bar, in the sense
that its stars have just enough energy (Jacobi's integral) to cross the
corotation resonance. 
\end{abstract}

\section{The Herculis stream}

\begin{figure}[t!]
\hspace*{.8cm}\psfig{file=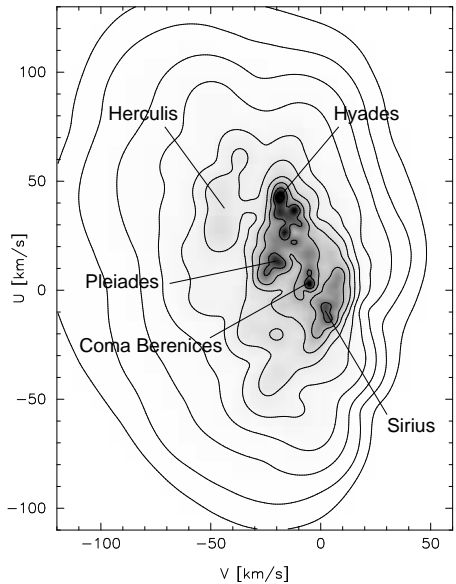,width=6cm}~\vspace*{-5.3cm}\\
\hspace*{6.8cm}\parbox{6.3cm}{
\caption{$U-V$ velocity distribution of the Hipparcos single stars
with distance $d<100$~pc, $\sigma\,(\pi)/\pi<0.1$ and radial velocities
in the input catalogue (3481 stars). The velocities are relative to the Sun;
$V$ is positive towards galactic rotation and $U$ towards the anti-centre.}
\vspace*{.8cm}
\label{obs}}
\end{figure}

Figure~\ref{obs} shows the observed velocity distribution in the $U$-$V$
plane, based on the Hipparcos input (radial velocities) and output (tangential
velocities) catalogues. The distribution is characterised by several distinct
streams, and in particular a stream with an asymmetric drift of
$40-50$~km\,s$^{-1}$ and $\overline{U}>0$, which hereafter we will refer to as
the ``Herculis'' stream according to a comoving Eggen group (Skuljan et al.
1999). This stream most probably has a dynamical origin because its stars are
mainly older than a few Gyr and present a large range of metallicities (Raboud
et al. 1998), and is under-represented in the Hipparcos sample, which is
biased towards young stars. The average luminosity of stars in this sample
indeed increases with distance and the sample only covers the vertical region
of the galactic plane where the fraction of young stars is largest.

\section{Simulation}

We have realised a large 3D $N$-body simulation designed to evolve into
realistic barred models of the Milky Way. The initial conditions are adapted
from Fux~(1997) and $N=1.4\times 10^7$ particles, allowing to compute very
detailed velocity distributions in small space regions.
\par After the formation of the bar (at $t\approx 1.4$~Gyr), the simulation
reveals a complex velocity structure with multiple streams occuring almost
everywhere in the disc. The streams, as well as the velocity dispersions,
remain very time-dependent until the end of the simulation ($t=2.65$~Gyr),
even within regions corotating with the bar. Outside corotation and for an
inclination angle of the bar $0<\varphi <90^{\circ}$, the simulation often
displays a Herculis-like stream (Figure~\ref{sim}).
The model selected here is a snapshot at $t=2.15$~Gyr, scaled to a
corotation radius of $(R_{12}+R_{45})/2=4.5$~kpc, where $R_{12}$ and $R_{45}$
are the galactocentric distances of the Lagrangian points $L_{\rm 1/2}$ and
$L_{\rm 4/5}$, and to a circular velocity in the azimuthally averaged potential
of $200$~km\,s$^{-1}$ at $R_{\circ}=8$~kpc. In these units,
$R_{\rm OLR}=7.7$~kpc and $\sigma_U:\sigma_V:\sigma_W=55:34:21$~km\,s$^{-1}$
at $R_{\circ}$ and $\varphi=25^{\circ}$.

\section{Interpretation}

\begin{figure}[t!]
\centerline{\psfig{file=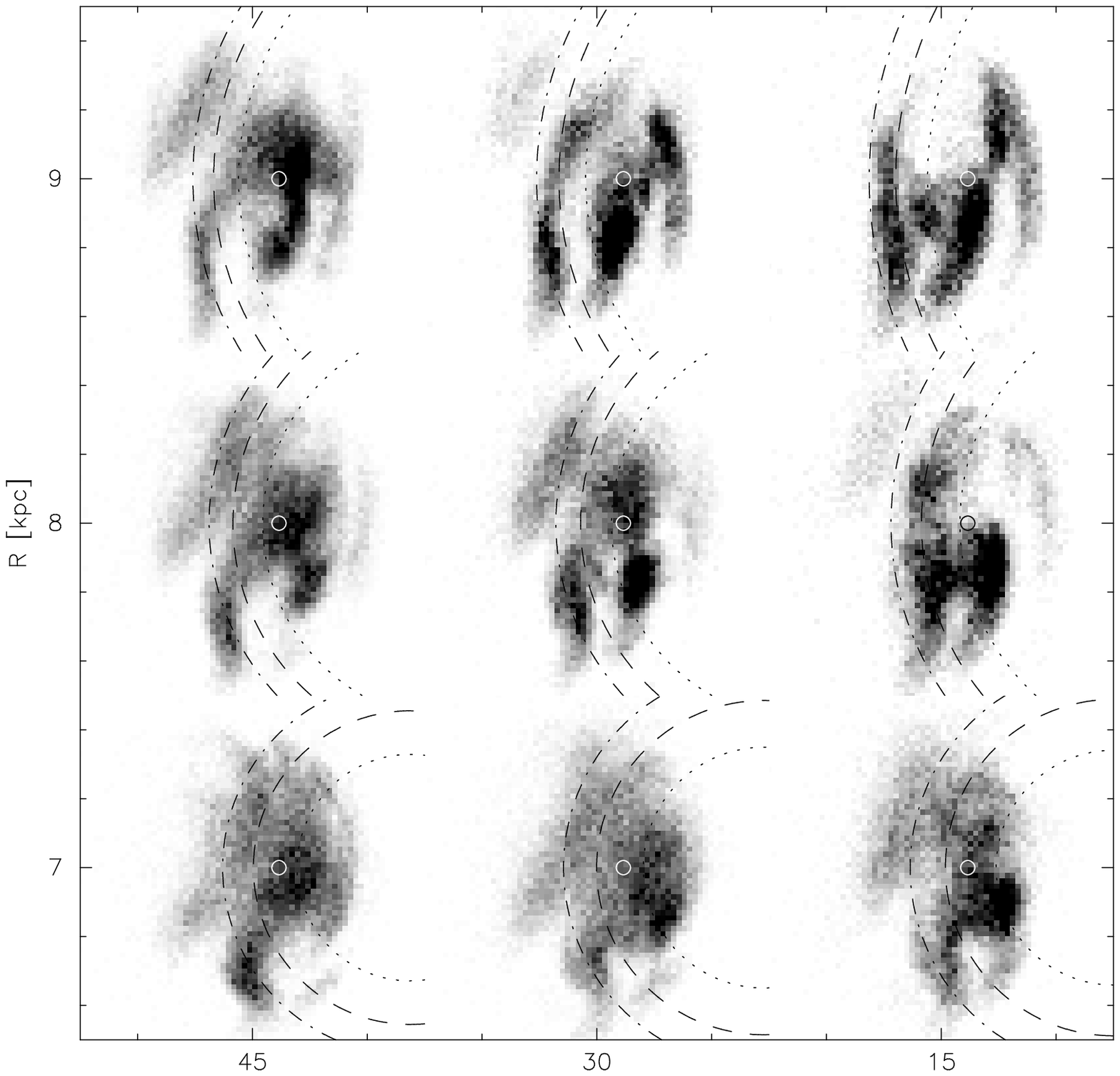,width=10.4cm}}
\centerline{\footnotesize \hspace*{.8cm}$\varphi \hspace*{.2cm} [^{\circ}]$}
\caption{Model $U-V$ distribution as a function of position in the galactic
disc ($t=2.15$~Gyr). The corotation is at $4.5$~kpc. The dashed-dotted, dashed
and dotted curves are the contours of constant Hamiltonian for $H=H_{45}$,
$H_{12}$ and $H_{\rm OLR}$ respectively (assuming~$W=0$). The circles indicate
the circular orbit in the axisymmetriesed potential.}
\label{sim}\vspace*{.5cm}
\centerline{\psfig{file=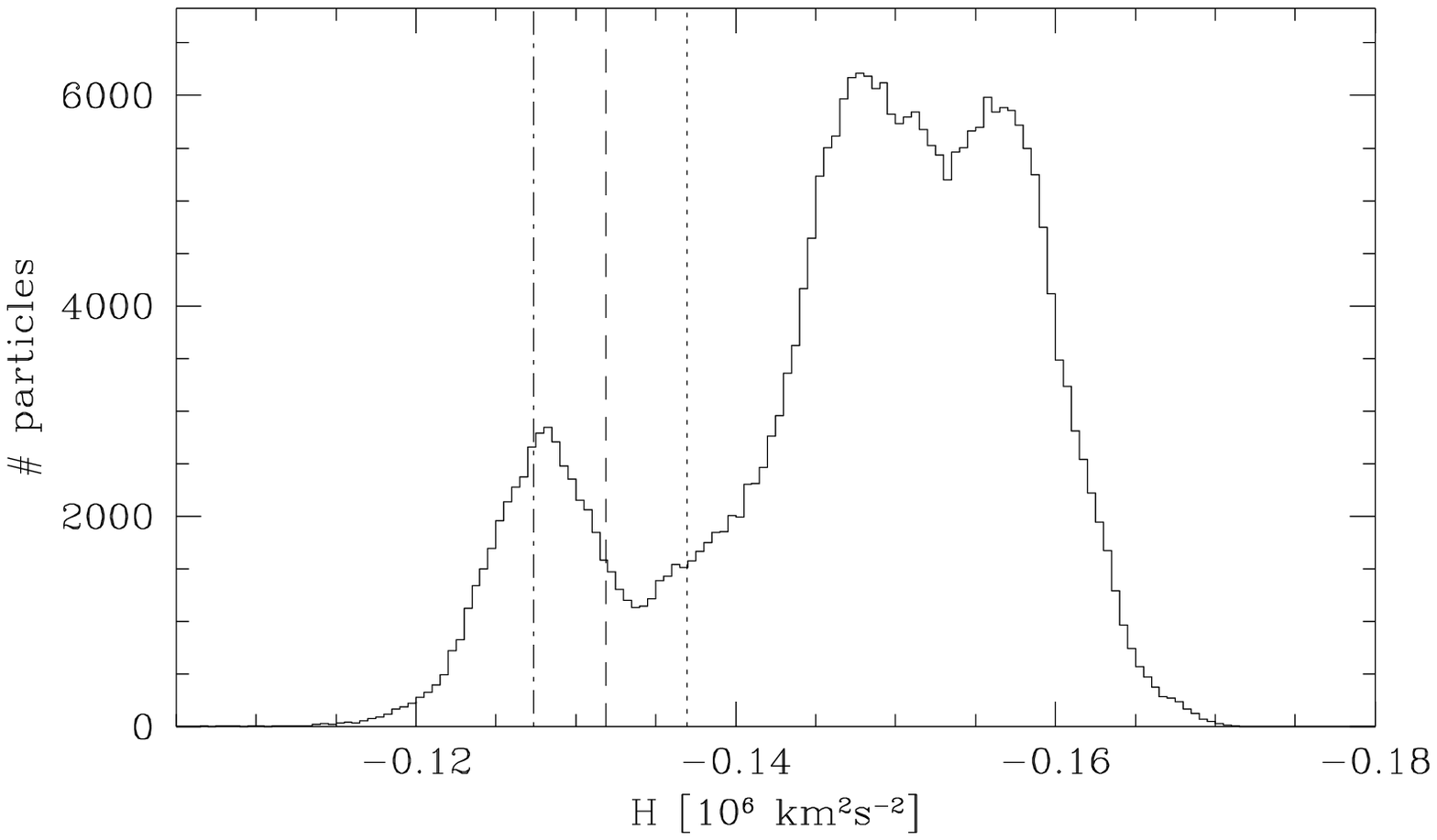,width=8.5cm}}
\caption{\hspace*{-.1cm}Hamiltonian distribution of the disc particles at
$R=11$~kpc ($t=2.15$~Gyr). The dashed-dotted, dashed and dotted vertical lines
respectively indicate the values $H_{45}$, $H_{12}$ and $H_{\rm OLR}$.
The part of the distribution with $H>H_{12}$ represents the hot particles
which may enter inside corotation.}\vspace*{-.7cm}
\label{hist}
\end{figure}

Our interpretation for the Herculis stream follows the same lines as in Raboud
et al. (1998). In a rotating barred potential, the value of the Hamiltonian:
\begin{equation}
H(\vec{x},\dot{\vec{x}})=\frac{1}{2}\dot{\vec{x}}^2+\Phi_{\rm eff}(\vec{x})
\label{ham}
\end{equation}
where $\Phi_{\rm eff}=\Phi(\vec{x})-\frac{1}{2}\Omega_{\rm P}(x^2+y^2)$,
$\Phi(\vec{x})$ is the gravitational potential and $\Omega_{\rm P}$ the
pattern speed of the bar, is an integral of motion (Jacobi's integral).
Three critical values of this integral are associated with stars corotating
at the Lagrangian points ($H_{12}\equiv \Phi_{\rm eff}(L_{1/2})$ and
$H_{45}\equiv \Phi_{\rm eff}(L_{4/5})$) and stars on circular orbits at
the outer Lindblad resonance in the axisymmetriesed potential ($H_{\rm OLR}$).
Stars with $H>H_{12}$ follow ``hot'' orbits which are susceptible to cross
the corotation radius. In the simulation, even far beyond the solar circle,
a substantial fraction of particles are on such orbits and the Hamiltonian
distribution frequently shows a secondary peak between $H_{12}$ and $H_{45}$
(Figure~\ref{hist}). If $U$ and $V$ are measured relative to the galactic
centre, Equation~(\ref{ham}) also writes:
\begin{equation}
(V-R\cdot \Omega_{\rm P})^2+U^2+W^2=2(H-\Phi_{\rm eff}).
\label{cont}
\end{equation}
Thus, neglecting the vertical velocity component $W$, the contours of constant
Hamiltonian in the $U-V$ plane are circles centred on
$(V,U)=(R\, \Omega_{\rm P},0)$ and of radius $2(H-\Phi_{\rm eff})$.
Figure~\ref{sim} shows that the Herculis-like stream in the model falls just
outside the $H_{12}$-contour and thus concerns particles on hot orbits.
\par This interpretation is different from the one proposed by Dehnen (1999),
who attributes the valley between the Herculis stream and the Hyades-Pleiades
branch to stars scattered off the OLR. In our case, the valley is produced by
the decline of the hot orbit population as $H\rightarrow H_{12}$. But in both
cases the position of the OLR is found near $R_{\circ}$, in agreement with
Kalnajs~(1991) conclusion based on the Hyades and Sirius streams.
\par
\begin{figure}[t!]
\centerline{\psfig{file=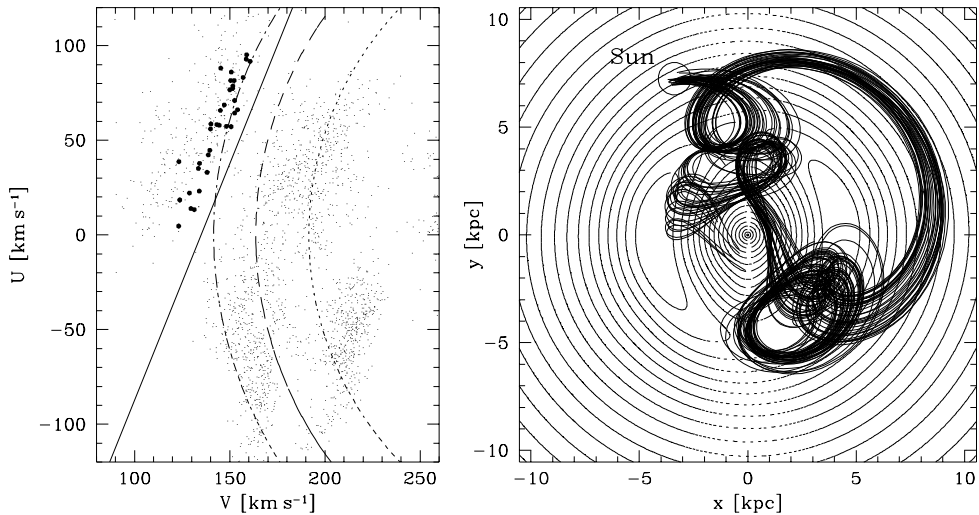,width=13cm}}
\caption{Left: model $U-V$ distribution for $R=8$~kpc and $\varphi=25^{\circ}$
($t=2.15$~Gyr). The circular contours are as in Figure~\ref{sim} and the bold
dots indicate the disc particles left of the diagonal line which stayed longer
than 300~Myr inside corotation during the past Gyr. Right: orbits of these
particles integrated 525~Myr backwards. The dotted lines are contours of
constant $\Phi_{\rm eff}$ and the pattern rotation is clockwise.}
\label{orb}
\end{figure}

 Figure~\ref{orb} traces back in time the orbits of the particles in the
Herculis-like stream which were recently captured inside corotation (using
the instantaneous frozen rotating potential).
These particles come from the low Hamiltonian edge of the stream and represent
about 10\% of the particles in the stream. This percentage strongly depends
on $R$ and $\varphi$.

\acknowledgments
The author thanks Ken Freeman and Agris Kalnajs for reading and commenting the
draft version of this paper.

\end{document}